\documentstyle[12pt,epsfig]{article}
\def\bge{\begin{equation}}
\def\ene{\end{equation}}
\def\bg{\begin{eqnarray}}
\def\en{\end{eqnarray}}

\def\D0bar{\overline{D^0}}

\begin{document}
%%%%%%%%%%%%%%%%%%%%%%%%%%%%%%%%%%%%%%%%%%%%%%%%%%%%%%%%%%%%%%%%%%%%%%%%%%%%%
\begin{titlepage}
\title{Effect of the bound nucleon form factors on 
charged-current neutrino-nucleus scattering}
\author{
K. Tsushima$^1$~\thanks{tsushima@ift.unesp.br}~\thanks{
Present address: IFT - UNESP,  
Rua Pamplona 145, 01405-900, S\~{a}o Paulo - SP, Brazil}, 
Hungchong Kim$^2$~\thanks{hung@phya.yonsei.ac.kr}~, and 
K. Saito$^3$~\thanks{ksaito@ph.noda.tus.ac.jp}
\\ \\
{$^1$\small Department of Physics and Astronomy,
University of Georgia, Athens, Georgia 30602, USA} \\
{$^2$\small Institute of Physics and Applied Physics, Yonsei 
University, Seoul, 120-749, Korea}\\
{$^3$\small 
Department of Physics, Faculty of Science and Technology}\\
{\small Tokyo University of Science, Noda 278-8510, Japan}
}
\date{}
\maketitle
\vspace{-9cm}
\hfill 
\vspace{9cm}
\begin{abstract}
We study the effect of bound nucleon form factors 
on charged-current neutrino-nucleus scattering.
The bound nucleon form factors of the 
vector and axial-vector currents are calculated in the 
quark-meson coupling model. 
We compute the inclusive 
$^{12}$C($\nu_\mu,\mu^-$)$X$ cross sections
using a relativistic Fermi gas model 
with the calculated bound nucleon form factors.
The effect of the bound nucleon form 
factors for this reaction is a reduction of $\sim$8\% for the 
total cross section, relative to
that calculated with the free nucleon form factors. 
\\ \\
%%% PACS 
%%%
%{\it Keywords}: Bound nucleon form factors, Neutrino-nucleus scattering,
%Relativistic Fermi gas model, Quark-meson coupling model
\end{abstract}
\end{titlepage}
%%%%%%%%%%%%%%%%%%%%%%%%%%%%%%%%%%%%%%%%%%%%%%%%%%%%%%%%%%%%%%%%%%%%%%%%%%%%%%

There has been of considerable interest in possible changes in the bound 
nucleon properties~\cite{change}.
A number of evidences, such as 
the nuclear EMC effect~\cite{EMC}, 
the quenching~\cite{gaspace,KTRiska} (enhancing~\cite{gatime}) 
of the space (time) component of the effective one-body axial coupling 
constant in nuclear $\beta$ decays,
the missing strength of the response functions in nuclear inelastic 
electron scattering and the suppression of the Coulomb sum 
rule~\cite{Coulomb}, have stimulated investigations of whether 
or not the quark degrees of freedom play any vital role. 

Recently, 
the electromagnetic form factors of 
bound protons were studied in polarized ($\vec e, e' \vec p$) scattering 
experiments on $^{16}$O and
$^4$He~\cite{O16He4}. The results from MAMI  
and Jefferson Lab on $^4$He~\cite{O16He4}
concluded that ratio of the electric ($G^p_E$)  
to magnetic ($G^p_M$) Sachs proton form factors
differs by 
$\sim$10\% in $^4$He from that in $^1$H.  
Conventional models 
employing free proton form factors,
phenomenological optical potentials, and bound state wave
functions, as well as relativistic corrections, 
meson exchange currents (MEC),
isobar contributions and final state 
interactions~\cite{O16He4,KELLY}, 
all fail to account for the observed effect in $^4$He~\cite{O16He4}.
Indeed, full agreement with the data was 
obtained only when, in addition
to these standard nuclear-structure corrections, 
a small correction due to 
the internal structure of the bound proton was taken into
account~\cite{O16He4,EMff}. 

Here, we study the effect of 
the bound nucleon form factors on neutrino-nucleus 
scattering~\footnote{Because the renormalization of axial-vector 
form factors are the 
same for the time and space components in this study (quenched), 
we will not discuss the time component.}.
As an example, we compute the inclusive 
$^{12}$C($\nu_\mu,\mu^-$)$X$ cross 
sections 
that have been measured by the LSND 
collaboration~\cite{LSND}.
It is known that 
the existing calculations
for the 
total cross section based on the nucleon and meson degrees
of freedom overestimate the data by  
$\sim$30\% to $\sim$100\%~\cite{LSND,Kolbe}. 
Because our aim is to focus on the effects 
due to the 
internal structure change of the bound nucleon, we use 
a relativistic Fermi gas model~\cite{Hungchong1,Hungchong2}, 
which is simple and transparent for the purpose,   
while implementing the bound nucleon form factors calculated 
in the quark-meson coupling (QMC) model~\cite{EMff,Axialff}. 
Thus, we do not include the other 
nuclear structure corrections~\cite{Kolbe,Delta}. 

Of course, it is difficult to  
separate exactly the effects we consider here from the standard 
nuclear-structure corrections, particularly from MEC.  
However, since the relevant current operators in this study 
are one-body quark (pion) operators 
acting on the quarks (pion cloud) 
in the nucleon, a double counting with the model-dependent 
MEC~\cite{KTRiska,KTRiska2} (the current operators act on the exchanged 
mesons) is expected to be avoided.
The same is also true for the model-independent meson pair currents,
because they are based on the {\it anti-nucleon} degrees of 
freedom~\cite{gaspace,KTRiska,KTRiska2}.
For the vector current, 
a double counting with MEC may be practically 
avoided because the analyses for the $^4$He($\vec{e},e'\vec{p}$)$^{\,3}$H 
experiments~\cite{O16He4} have shown. 
For the 
axial-vector current, the quenching of the axial 
coupling constant 
($g_A = G_A(0)$) due to the model-independent meson pair currents 
was estimated~\cite{KTRiska} using a Fermi gas model.
The quenching due to the pair currents 
amounts to a 2\% at normal 
nuclear matter density, thus contributing negligibly
to the cross section.
Hence, the double counting from the interference 
between the axial-vector and vector currents  
is also expected to be small, in considering    
the analyses for the $^4$He($\vec{e},e'\vec{p}$)$^{\,3}$H 
experiments. 
Thus, the effect we consider here, which originates from the change of 
the internal quark wave function, is additional 
to the standard nuclear-structure corrections.
 
The QMC model~\cite{qmc} 
has been successfully applied to many problems of 
nuclear physics and hadronic properties in nuclear medium~\cite{QMCapp}.
In the model, the medium effects arise through 
the self-consistent coupling of scalar ($\sigma$) and vector 
($\omega$) meson fields to confined quarks, rather than to the nucleons. 
As a result, the internal structure of the bound nucleon is modified by the 
surrounding nuclear medium. 
(Details of the QMC model are given in Refs.~\cite{qmc,QMCapp}.)

Assuming G-parity (no second-class current), the charged-current vector 
and axial form factors for free nucleons with mass $m_N$ are defined by:
\bg
\left< {p' s'}| V^\mu_a(0) |{p s} \right> 
&=& \overline{u}_{s'}(p')\left[F_1(Q^2)\gamma^\mu +
i (F_2(Q^2) / 2m_N) \sigma^{\mu \nu} (p'-p)_\nu\right] 
(\tau_a / 2) u_s(p),
\label{vff}\\
\left< {p' s'}| A^\mu_a(0) |{p s} \right> 
&=& \overline{u}_{s'}(p')\left[G_A(Q^2)\gamma^\mu +
 (G_P(Q^2) / 2m_N) (p'-p)^\mu\right]\gamma_5 (\tau_a / 2) u_s(p),
\label{aff}
\en
where $Q^2 \equiv -(p'-p)^2$, and other notations should be selfexplanatory.
The vector form factors, $F_1(Q^2)$ and $F_2(Q^2)$, 
are related to the electric ($G_E(Q^2)$) and 
magnetic ($G_M(Q^2)$) Sachs form factors by 
the conserved vector current hypothesis. 
The induced pseudoscalar form factor, $G_P(Q^2)$, 
is dominated by the pion pole and can be calculated using the 
PCAC relation~\cite{CBM}. 
Nevertheless, the contribution from $G_P(Q^2)$ to the cross section 
is proportional to (lepton mass)$^2/m_N^2$, and small
in the present study.  
We note that, since there is another vector in nuclear medium, 
the nuclear (matter) four-velocity,
there may arise various other form factors 
in addition to those in Eqs.~(\ref{vff}) and~(\ref{aff}).
The modification of the nucleon internal structure
studied here may also be expected to contribute to such form factors. 
However, at this stage, information on such form factors 
is very limited and not well under control  
in theoretically and experimentally. 
Thus, we focus on the in-medium changes of the free form 
factors given in Eqs.~(\ref{vff}) and~(\ref{aff}), and 
study their effects on neutrino-nucleus scattering.
(Hereafter we denote the in-medium quantities by an asterisk $^*$.)

Using the improved cloudy bag model (ICBM)~\cite{ICBM} and QMC, the 
electromagnetic and axial form factors in nuclear medium 
are calculated in the Breit frame~\cite{EMff,Axialff}:
\bg
G_{E,M,A}^{QMC\, *}(Q^2) &=& \eta^2\ G^{\rm sph\, *}_{E,M,A}(\eta^2 Q^2)\ ,
\label{GEMA}
\en
where $\eta = (m_N^*/E_N^*)$ is the scaling factor 
with $E_N^*=\sqrt{{m_N^*}^2 + Q^2/4}$ the energy, 
and $m_N^*$ the effective nucleon mass in nuclear medium.
The explicit expressions for Eq.~(\ref{GEMA}) 
are given in Refs.~\cite{EMff,Axialff}.
The ICBM includes a Peierls-Thouless projection to account for
center of mass and recoil corrections, and a Lorentz contraction of the
internal quark wave function~\cite{ICBM,Lorentz}. 

Now we calculate the ratios of the bound to free nucleon form factors, 
[$G_{E,M,A}^{QMC\,*}/G_{E,M,A}^{ICBM\,free}$], 
to estimate the bound nucleon form factors. 
Using the empirical parameterizations 
in free space $G_{E,M,A}^{emp}$~\cite{EMffpara,Axialffpara}, 
the bound nucleon form factors $G_{E,M,A}^*$ are calculated by 
\bge
G_{E,M,A}^*(Q^2) = 
\left[ G_{E,M,A}^{QMC*}(Q^2) / G_{E,M,A}^{ICBM\, free}(Q^2) \right]
G_{E,M,A}^{emp}(Q^2).
\label{boundNff}
\ene
Note that the pion cloud effect is not included in the axial 
form factor 
in the present treatment~\cite{Axialff}. 
However, the normalized $Q^2$ dependence (divided by $g_A = G_A(0)$) 
relatively well reproduces the empirical
parameterization~\cite{Axialff}. 
Furthermore, the relative modification of $G^*_A(Q^2)$  
due to the pion cloud is expected to be small, 
since the pion cloud contribution 
to entire $g_A$ is $\sim$8\%~\cite{CBM} without a specific 
center-of-mass correction.

In the calculation we use the parameter values, the current quark mass 
$m_q (= m_u = m_d) = 5$ MeV assuming SU(2) symmetry, 
and the free nucleon bag radius   
$R_N = 0.8$~fm, where both values are considered to be standard 
in QMC~\cite{qmc}. 

\begin{figure}[hbt]
\begin{center}
\epsfig{file=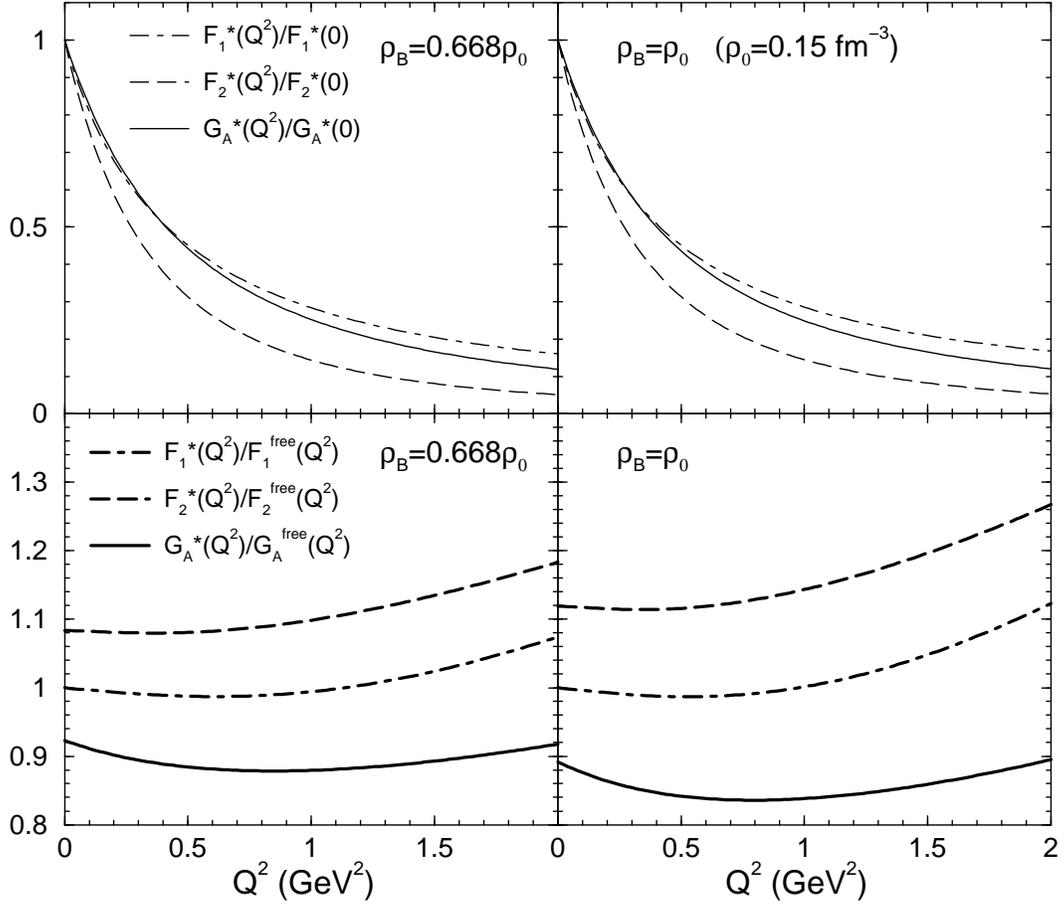,height=14cm,angle=-90}
\caption{Calculated ratios for the bound nucleon form factors.
\label{mediumff}
}
\end{center}
\end{figure}

First, 
Fig.~\ref{mediumff} shows 
ratios of the bound to free nucleon form factors calculated 
as a function of $Q^2$ 
for $\rho_B = \rho_0 = 0.15$ fm$^{-3}$ (the normal nuclear matter density) and 
$0.668 \rho_0$ (the Fermi momentum $k_F = 225$ MeV for $^{12}$C).
The lower panels in Fig.~\ref{mediumff} show the enhancement of 
momentum dependence of $F_2^*(Q^2)$ and 
$G_A^*(Q^2)$, as well as the enhancement of $F^*_2(0)$ and 
quenching of $G^*_A(0)$~\cite{EMff,Axialff,Saitoff}. 
Although the modification of the $Q^2$ dependence  
is small, we emphasize that this effect
originates from the nucleon internal structure change. 
The main origin of this new $Q^2$ dependence is the
Lorentz contraction effect to the quark wave function 
amplified by the reduced effective nucleon 
mass. (See also Eq.~(\ref{GEMA}).)
Note that, the relative change of the bound nucleon form factor 
$F_2^*(Q^2) [G_E^{p*}(Q^2)]$ to that of the free nucleon 
is an enhancement [quenching~\cite{EMff}] of 
$\sim$8\% [4\%] in $^{12}$C at $Q^2 = 0.15$ GeV$^2$, and we are focusing 
on such relative change.

Next, we 
investigate the effect of the bound nucleon form factors 
on charged-current neutrino-nucleus scattering. 
We compute 
the inclusive $^{12}$C($\nu_\mu,\mu^-$)$X$ differential and total 
cross sections, which have been measured 
by the LSND collaboration~\cite{LSND}.
We use the formalism described in Ref.~\cite{Hungchong1}, and that the 
empirical parameterizations of the electromagnetic~\cite{Hungchong1,EMffpara} 
and axial~\cite{Axialff,Axialffpara} form factors 
for the free nucleon. 
(See Eq.~(\ref{boundNff}).) 
A relativistic Fermi gas model is used implementing the bound nucleon
form factors to calculate the
differential cross section 
$\left< d\sigma/dE_\mu \right>$, averaged over the LSND muon neutrino 
spectrum $\Phi(E_{\nu_\mu})$~\cite{Hungchong1} 
for the full range of the LSND experimental spectrum~\cite{LSND}, 
$0 \le E_{\nu_\mu} \le 300$ MeV:
\bge
\left< d\sigma / dE_\mu \right> = 
\left[ \int_0^\infty (d\sigma/dE_\mu) \Phi(E_{\nu_\mu}) dE_{\nu_\mu} \right]
/ 
\left[ \int_0^\infty \Phi(E_{\nu_\mu}) dE_{\nu_\mu} \right]\ .
\label{dsigma}
\ene

\begin{figure}[hbt]
\begin{center}
\epsfig{file=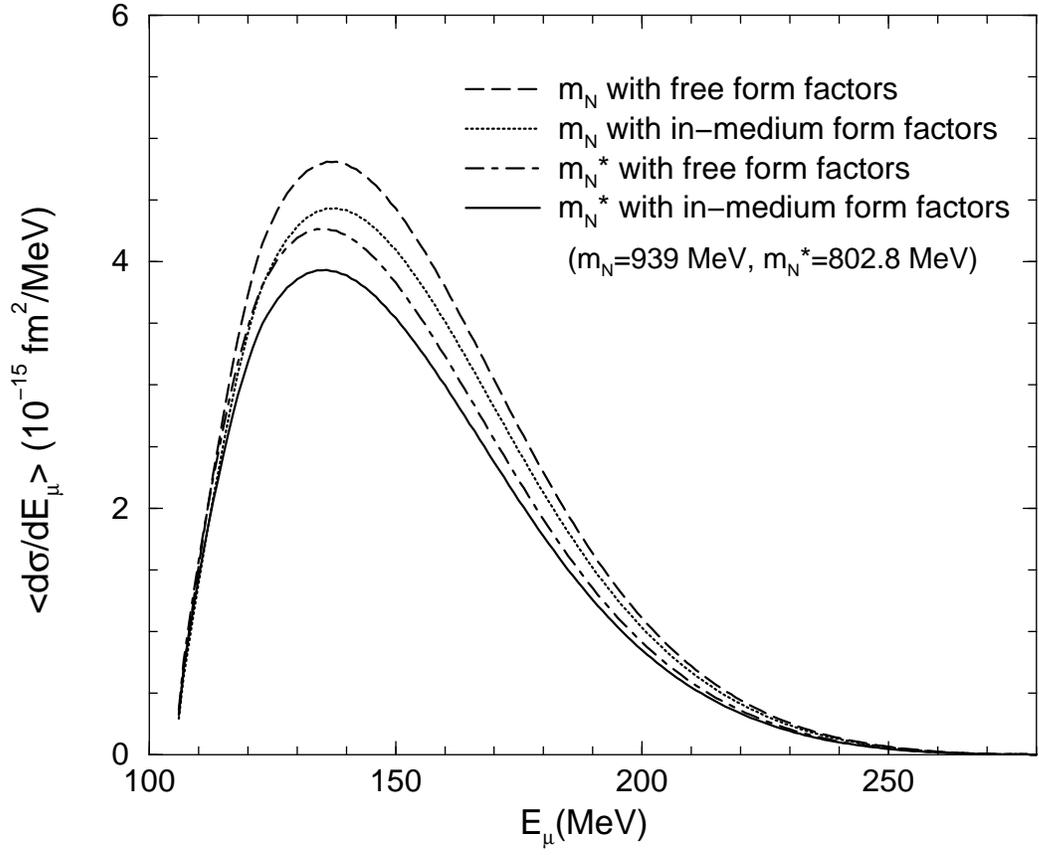,height=14cm,angle=-90}
\caption{Angle-integrated inclusive 
$^{12}$C($\nu_\mu,\mu^-$)$X$ differential cross section as 
a function of the emitted 
muon energy $E_\mu$ using $E_B = 0$ for all cases.
\label{QMCdsigma}
}
\end{center}
\end{figure}

\begin{figure}[hbt]
\begin{center}
\epsfig{file=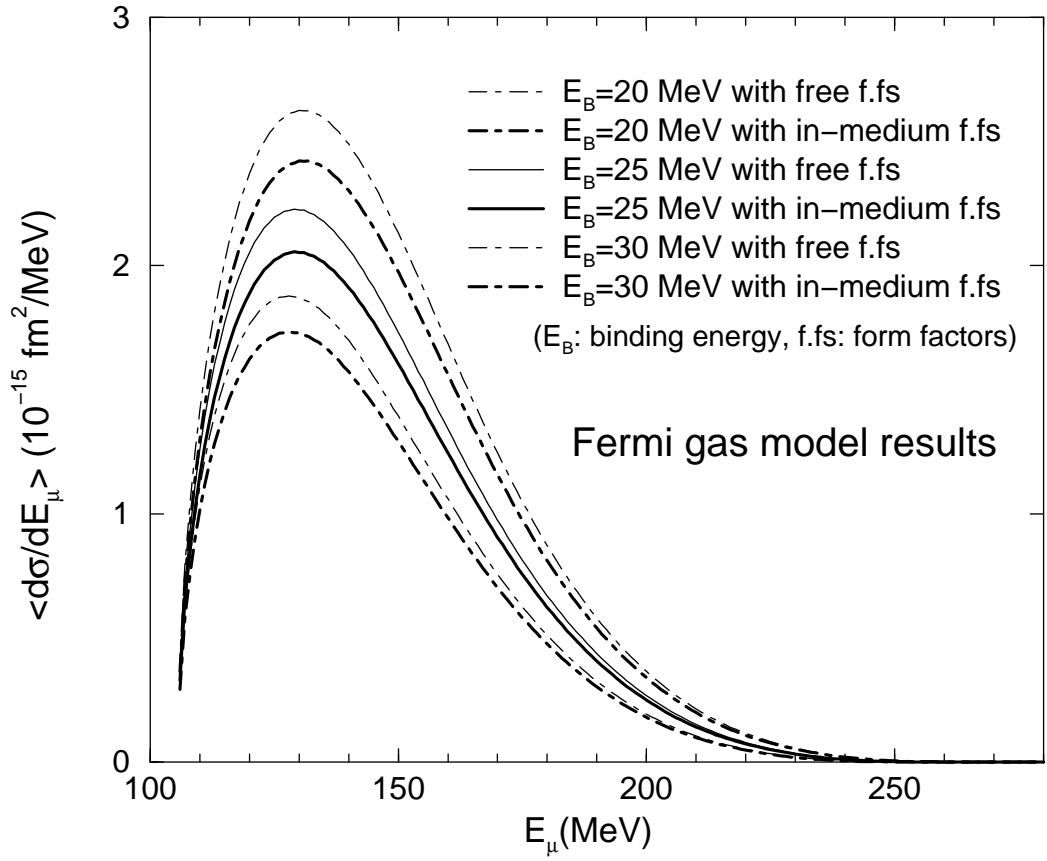,height=14cm,angle=-90}
\caption{Same as Fig.~\protect\ref{QMCdsigma}, 
but using $m_N = 939$ MeV for all cases. 
\label{Fermigas}
}
\end{center}
\end{figure}

Fig.~\ref{QMCdsigma} shows the result of 
$\left< d\sigma / dE_\mu \right>$ calculated using the nucleon masses, 
$m_N$ and $m_N^*$.
For the Fermi momentum $k_F = 225$ MeV ($\rho_B = 0.668 \rho_0$) for 
$^{12}$C, we use the QMC calculated value, $m_N^* = 802.8$ MeV.
A moderate quenching of the cross section can be observed
due to the in-medium form factors for both cases. 
Although the effective nucleon mass 
can account for, to some extent, the binding  
effect (the Hugenholtz-van Hove theorem~\cite{binding}), 
there is an alternative to 
include the binding effect, i.e., 
the ``binding energy'' $E_B$ is introduced and the available reaction energy 
$E$ is replaced by $E - E_B$. 
In this case, we use the free nucleon mass in the calculation.
Since $E_B$ is an effective way of accounting for 
the binding effect~\cite{eb}, we regard $E_B$ as a parameter and 
perform calculations for $E_B = 20, 25$ and $30$ MeV.
(E.g., $E_B = 25-27$ MeV is commonly used 
for the $^{16}$O nucleus~\cite{be25}.)
We emphasize that our aim is not to reproduce the LSND 
data, but to estimate the corrections due to the 
bound nucleon form factors.
In Fig.~\ref{Fermigas} we present the 
results of $\left< d\sigma / dE_\mu \right>$ for 
$E_B = 20, 25$ and $30$ MeV. 
In both Figs.~\ref{QMCdsigma} and~\ref{Fermigas}, 
the bound nucleon form factors reduce 
the differential cross section. 
In Fig.~\ref{Fermigas}, as the binding energy $E_B$ increases, the peak 
position shifts downward for both cases  
with the free and bound nucleon form factors. 
The similar tendency due to $m^*_N$ 
is also seen in Fig.~\ref{QMCdsigma}. 

The total cross section is given by integrating Eq.~(\ref{dsigma}) 
over the muon energy. 
We denote the cross section calculated with the free [bound] nucleon 
form factors, $F_{1,2}(Q^2)$ and $G_{A,P}(Q^2)$
[$F^*_{1,2}(Q^2)$ and $G^*_{A,P}(Q^2)$], 
as $\left<\sigma(F,G)\right>$ [$\left<\sigma(F^*,G^*)\right>$]. 
Thus, $\left<\sigma(F,G)\right>$ calculated with 
$m_N$ and $E_B = 0$ corresponds to the free Fermi gas model result.  
The results 
with $E_B = 0$ and either $m_N$ or $m_N^*$ are listed in the 
top group rows in Table~\ref{table}.
The LSND experimental data~\cite{LSND} are also shown in the bottom 
group rows in Table~\ref{table}.
As expected~\cite{LSND}, the free Fermi gas result
overestimates the data 
by a factor of three. The results obtained using the bound nucleon 
form factors, 
with either $m_N$ or $m_N^*$, similarly overestimate the LSND data.
In order to make discussions more quantitative, we define:
\bge
R(\delta\sigma) \equiv \left[ \left<\sigma(F,G)\right> - 
\left<\sigma(F^*,G^*)\right> \right] / \left<\sigma(F,G)\right>. 
\label{delsigma}
\ene
For the total cross sections calculated with $(m_N, m_N^*)$ and $E_B = 0$, 
we get $R(\delta\sigma) = (7.7, 7.7)$\%, respectively. 
Thus, the correction due to the bound nucleon form factors 
to the total cross section 
is not sensitive to $m_N$ or $m_N^*$ in the case of $E_B = 0$. 

Next, we investigate which bound nucleon form factor 
gives dominant corrections to the total cross section. 
We calculate the total cross section with $m_N$, 
using the free and bound form factors for two cases,  
[$F^*_{1,2}(Q^2)$ and $G_{A,P}(Q^2)$] and 
[$F_{1,2}(Q^2)$ and $G^*_{A,P}(Q^2)$]. 
They 
are denoted by 
$\left<\sigma(F^*,G)\right>$ and $\left<\sigma(F,G^*)\right>$, respectively.
The results are given in the middle group rows in Table~\ref{table}. 
Together with the results in the upper group rows in 
Table~\ref{table}, we obtain inequalities for the total cross 
sections calculated with $m_N$ and $E_B = 0$:
\bge
\left<\sigma(F,G^*)\right> < \left<\sigma(F^*,G^*)\right> 
< \left<\sigma(F,G)\right> < \left<\sigma(F^*,G)\right>.
\label{inequality}
\ene
This shows that the most dominant reduction
is driven by the axial form factor,
$G_A^*(Q^2)$. 
(The induced pseudoscalar form factor 
$G_P(Q^2)$  
gives only a few percent contribution  
when calculated using all free form factors.)
Furthermore, 
$F^*_{1,2}(Q^2)$ enhance the total cross section (mostly due to $F^*_2(Q^2)$)
as can be 
seen from the lower panel in Fig.~\ref{mediumff}. 

The total cross sections 
for $E_B = 20, 25$ and $30$ MeV  
are listed in the bottom group rows in Table~\ref{table}. 
The bound nucleon form factors for these cases also 
reduce the total cross section relative to those calculated 
with the free form factors.
In addition, the results are rather sensitive 
to the values for $E_B$.
However, we find    
$R(\delta\sigma) = (8.1,7.6,7.5)\%$ for $E_B = (20,25,30)$MeV, 
respectively. Thus, the effect of the bound nucleon form factors 
to the reduction rate is again not sensitive to $E_B$. 

To summarize, we have estimated the effect of the bound nucleon
form factors arising from the nucleon internal structure change
on the inclusive $^{12}$C($\nu_\mu,\mu^-$)$X$  
cross sections. 
We have used a relativistic Fermi gas model implementing
the bound nucleon form factors calculated in the QMC model.
The effect of the bound nucleon form factors for this reaction 
is a reduction of $\sim$8\% for the
total cross section. This $\sim$8\% reduction
(or an order of 10\% for a heavier nucleus)
should be taken into account additionally to
the standard nuclear-structure corrections.
To draw a more definite conclusion, it is essential to perform a more
precise, elaborate calculation within the framework of
RPA~\cite{Hungchong1}
including the effect of bound nucleon form factors. 
However, even at the present stage,
it is important to point out that
the correction due to the in-medium form factors
could be significant for a precise estimate
of the charged-current neutrino-nucleus cross section.

\vspace{1ex}
\noindent
{\bf Acknowledgments}\\
We would like to thank D.H. Lu for providing us the improved cloudy bag 
model codes, and 
K. Kubodera, K. Nakayama, and R. Seki 
for useful discussions.
K.T. was supported by the Forschungszentrum-J\"{u}lich, 
contract No. 41445282 (COSY-058). 

%\newpage
%%%%%%%%%%%%%%%%%%%%%%%%%%%%%%%%%%%%%%%%%%%%%%%%%%%%%%%%%%%%%%%%%%%%%%%%%%

%%%%%%%%%%%%%%%%%%%%%%%%%%
%%%%%%%% Table
%%%%%%%%%%%%%%%%%%%%%%%%%%
\newpage
%%%%%%%%%%%%%%%%%%%%%%%%%%%%%%%%%%%%%%%%%%%%%%%%%%%%
\begin{table}[hbtp]
\begin{center}
\caption{Calculated 
total cross sections for 
$^{12}$C($\nu_\mu,\mu^-$)$X$. 
See the text for notations. 
}
\vspace{3ex}
\begin{tabular}{llcc}
\hline\hline
Notation& Type of calculation 
&$E_B$ (MeV) 
&$\left<\sigma\right>$ in 10$^{-40}$ cm$^2$\\
\hline\hline
$\left<\sigma(F,G)\right>$ &$m_N,F_{1,2}(Q^2),G_{A,P}(Q^2)$ &0 &32.5\\ 
$\left<\sigma(F^*,G^*)\right>$ &$m_N,F_{1,2}^*(Q^2),G_{A,P}^*(Q^2)$ &0 &30.0\\ 
$\left<\sigma(F,G)\right>$ &$m_N^*, F_{1,2}(Q^2),G_{A,P}(Q^2)$ &0 &28.4\\ 
$\left<\sigma(F^*,G^*)\right>$ &$m_N^*,F_{1,2}^*(Q^2),G_{A,P}^*(Q^2)$&0&26.2\\ 
\hline\hline
$\left<\sigma(F^*,G)\right>$& $m_N, F_{1,2}^*(Q^2), G_{A,P}(Q^2)$ &0   &33.5\\ 
$\left<\sigma(F,G^*)\right>$& $m_N, F_{1,2}(Q^2), G_{A,P}^*(Q^2)$ &0   &29.1\\ 
\hline\hline
$\left<\sigma(F,G)\right>$& $m_N, F_{1,2}(Q^2), G_{A,P}(Q^2)$      &20  &16.1\\ 
$\left<\sigma(F^*,G^*)\right>$& $m_N,F_{1,2}^*(Q^2),G_{A,P}^*(Q^2)$&20  &14.8\\ 
$\left<\sigma(F,G)\right>$& $m_N,F_{1,2}(Q^2),G_{A,P}(Q^2)$        &25  &13.2\\ 
$\left<\sigma(F^*,G^*)\right>$& $m_N,F_{1,2}^*(Q^2),G_{A,P}^*(Q^2)$&25  &12.2\\ 
$\left<\sigma(F,G)\right>$& $m_N,F_{1,2}(Q^2),G_{A,P}(Q^2)$        &30  &10.7\\ 
$\left<\sigma(F^*,G^*)\right>$& $m_N,F_{1,2}^*(Q^2),G_{A,P}^*(Q^2)$&30  & 9.9\\ 
\\
 &Experiment~\protect\cite{LSND} (2002)& &$10.6 \pm 0.3 \pm 1.8$\\
 &Experiment~\protect\cite{LSND} (1997)& &$11.2 \pm 0.3 \pm 1.8$\\
 &Experiment~\protect\cite{LSND} (1995)& &$ 8.3 \pm 0.7 \pm 1.6$\\
\hline\hline
\end{tabular}
\label{table}
\end{center}
\end{table}
%%%%%%%%%%%%%%%%%%%%%%%%%%%%%%%%%%%%%%%%%%%%%%%%%%%%
 
%%%%%%%%%%%%%%%%%%%%%%%%%%
%%%%%%%% Figures
%%%%%%%%%%%%%%%%%%%%%%%%%%
\newpage
%%%%%%%%%%%%%%%%%%%%%%%%%%%%%%%%%%%%%%%%%%%%%%%%%%%%
%%%%%%%%%%%%%%%%%%%%%%%%%%%%%%%%%%%%%%%%%%%%%%%%%%%%
\newpage
%%%%%%%%%%%%%%%%%%%%%%%%%%%%%%%%%%%%%%%%%%%%%%%%%%
%%%%%%%%%%%%%%%%%%%%%%%%%%%%%%%%%%%%%%%%%%%%%%%%%%%%
\newpage
%%%%%%%%%%%%%%%%%%%%%%%%%%%%%%%%%%%%%%%%%%%%%%%%%%
%%%%%%%%%%%%%%%%%%%%%%%%%%%%%%%%%%%%%%%%%%%%%%%%%%%%

%
%%%%%%%%%%%%%%%%%%%%%%
\end{document}